\documentclass[aps,epsfig,twocolumn,showpacs,superscriptaddress,showkeys]{revtex4}
\usepackage{amsmath}
\usepackage{epsfig}
\usepackage{graphics}
\def\vec#1{\mathbf{#1}}
\begin{document}
\title{Multi-channel scattering and Feshbach resonances:\\
 Effective theory, phenomenology,  and many-body effects}
\author{G.M.~Bruun}
\affiliation{The Niels Bohr Institute, Blegdamsvej 17, DK-2100
Copenhagen, Denmark}
\author{A.D.~Jackson}
\affiliation{The Niels Bohr Institute, Blegdamsvej 17, DK-2100
Copenhagen, Denmark}
\author{E.E.~Kolomeitsev}
\affiliation{The Niels Bohr Institute, Blegdamsvej 17, DK-2100
Copenhagen, Denmark}
\affiliation{School of Physics and Astrophysics, University of Minnesota, Minneapolis,
MN 55455, USA}

\begin{abstract}
A low energy effective theory based on a microscopic multi-channel
description of the atom-atom interaction is derived for the scattering
of alkali atoms in different hyperfine states.  This theory
describes all scattering properties, including medium effects, in
terms of the singlet and triplet scattering lengths and the range
of the atom-atom potential and provides a link between a microscopic
description of Feshbach scattering and more phenomenological approaches.
It permits the calculation of medium effects on the
resonance coming from the occupation of closed channel states. The examination of such effects  are
demonstrated to be of particular relevance to an experimentally important
Feshbach resonance for $^{40}$K atoms.  We analyze a recent
rethermalization rate experiment on $^{40}$K and demonstrate that a measurement of the temperature dependence 
of this rate can determine the magnetic moment of the Feshbach
molecule.  Finally, the energy dependence of the Feshbach interaction is
shown to introduce a negative effective range inversely proportional
to the width of the resonance. Since our theory is based on a microscopic
multi-channel picture, it allows the explicit calculation of corrections
to commonly used approximations such as the neglect of the effective range
and the treatment of the Feshbach molecule as a point boson.
\end{abstract}
\date{\today}
\maketitle

\section{Introduction}
The use of Feshbach resonances to vary the interactions between atoms
obeying Fermi statistics has become a matter of interest in the field of
ultracold trapped atoms~\cite{Experiments}.  One reason
is that these systems permit the experimental exploration of potentially
strongly correlated Fermi systems in which the magnitude of the scattering
length $a$ is larger than the interparticle separation. The study of such
strongly interacting Fermi systems is relevant to several fields including
condensed matter and nuclear physics and astrophysics.

It is therefore
important to understand in some detail the physics of atoms interacting via
a Feshbach resonance.
Theoretical studies of the two-body physics of Feshbach resonances are
typically performed in the context of vacuum multi-channel scattering theory
using a microscopic model of the atom-atom interaction~\cite{2body}.  On the
other hand, studies of low energy many-body properties often require
the use of effective theories with a simplified form of the atom-atom
interaction in order to make calculations tractable.  A precise link between
such effective theories and microscopic multi-channel models is therefore
crucial in order to incorporate the correct two-body physics in many-body
calculations.  In the atomic gas community, effective theories are of two
types.  In one, the presence of closed channels is simply neglected, and
only the resultant (resonant) scattering length is used as an input parameter.
This approximation is adequate for the calculation of certain many-body
properties close to broad resonances where the wave function of the Feshbach
molecule is dominated by atom pairs in the open channel~\cite{Randeria}.
In the other, the presence of a two-body bound state in the closed channel
(which leads to the resonance) is put in by hand in the form of a point
boson~\cite{Feshbachrefs,bruunpethick}.  The parameters appearing in such
theories are then the magnetic moment of the Feshbach molecule  and the
molecule-atom coupling.

The treatment of the bound state in the closed channels as a point boson is,
however, incorrect in principle since it is really a composite two-body
object. The main purpose of the present paper is therefore to develop a
systematic, improved low energy effective theory for the multi-channel
scattering of alkali atoms in which the presence of a closed channel
molecular Feshbach state is not put in by hand as a point boson but rather
emerges naturally as  a composite object consisting of two interacting
fermionic atoms.  We will show that such a theory permits tractable
calculations of low energy many-body properties including medium effects
due to the occupation of states in the closed channels.  The examination of such effects will
be demonstrated to be of particular importance for the experimentally
well-studied resonance of $^{40}$K atoms.  It will be demonstrated that
our approach can be reduced to the well-known Bose-Fermi model when the
two-fermion nature of the bare molecule in the closed channel can be ignored.
This theory can be formulated in two equivalent ways: One uses the microscopic
parameters describing the atom-atom interaction as input; the other takes
physical parameters describing the low energy scattering as input
including the position and width of the resonance and the magnetic moment
of the Feshbach molecule (including the effects of dressing by the open
channel).  This provides an appealing link between the microscopic parameters
often used in two-body multi-channel calculations and the physical
parameters used in low energy effective many-body theories.  We will
use this effective theory to analyze recent rethermalization experiments
on $^{40}$K atoms and propose that further experimental study of
rethermalization rates at higher temperatures can provide a determination
of the magnetic moment of the Feshbach molecule.  Finally, the energy
dependent Feshbach interaction will be shown to introduce an effective range
inversely proportional to the width of the resonance.  This effective range
can be neglected (and universal behavior close to resonance expected) only
in the case of broad resonances.

Note that we set $\hbar=1$ throughout the paper.

\section{Low energy effective theory}
In this section, we develop a low-energy effective theory for the scattering
of alkali atoms in different hyperfine states.  Consider a gas of alkali atoms
in a given magnetic field $B$ along the $z$-direction. Incoming and outgoing
scattering channels are characterized by the eigenstates of the single
particle hyperfine Hamiltonian
\begin{equation}
\hat{H}_{\rm spin}=A\, \vec{I}\cdot \vec{S}+C\, S_z+D\, I_z
\label{Hspin}
\end{equation}
where $A$ is the hyperfine constant, $C=2\mu_B\,B$, $D=-\mu_n\,B/I$ with
$\mu_n$ the nuclear spin magnetic moment, and ${\vec{S}}$ and ${\vec{I}}$
are the electron and nuclear spin operators respectively.  We denote the
single particle eigenstates of the hyperfine Hamiltonian for a given
$B$-field as $|\alpha\rangle$ with $\hat{H}_{\rm spin} |\alpha\rangle
= \epsilon_\alpha|\alpha\rangle$.  The eigenstates can be labeled by their
total spin, $\vec{F}=\vec{S}+\vec{I}$, $F=I\pm\frac12$, and its $z$-projection,
$m_F=-I-\frac12\,,\dots I+\frac12$, e.g.,  $|\alpha\rangle \equiv
|F,m_F\rangle$.  The main contribution to the atom-atom interaction
driving the scattering process is the electrostatic central potential
\begin{equation}
V(r)=\frac{V_s(r)+3\,V_t(r)}{4}+[V_t(r)-V_s(r)]\,{\vec{S}}_1\cdot{\vec{S}}_2
\label{vcentral}
\end{equation}
where $V_s(r)$ and $V_t(r)$ are the singlet and triplet potentials
and ${\mathbf{S}}_1$ and ${\mathbf{S}}_2$ are the spins of the valence
electrons of the two alkali atoms~\cite{PethickBook}. We ignore
here the small magnetostatic and magnetic dipole-dipole interactions between
atoms.  The potential (\ref{vcentral}) induces transitions among
two-particle states with the same total $z$-projection of $\vec{F}$\,.
The properly anti-symmetrized two-particle state with the lowest energy
constitutes the open channel $|o\rangle=|\alpha_1,\alpha_2\rangle$.
The states of higher energies which can be coupled to $|o\rangle$ by
the potential (\ref{vcentral}) form a set of closed channels
$|c^{(n)}\rangle=|\alpha_3^{(n)},\alpha_4^{(n)}\rangle$.  The threshold
energies for the closed channels are then  $E^{(n)}_{\rm th}(B) =
\epsilon_{\alpha_4^{(n)}}+\epsilon_{\alpha_3^{(n)}} - \epsilon_{\alpha_2}
-\epsilon_{\alpha_1}$, and they depend on the magnetic field $B$ through
$\hat{H}_{\rm spin}$.

As a practical illustration of our approach, we consider in the following
the s-wave Feshbach resonance in a cold $^{40}K$ gas at
$B_0 \simeq 201.6$~G, which has frequently been studied experimentally,
e.g., Refs.~\cite{Loftus,Regal}.  The open channel for this resonance
has atoms in the states $|\frac92,-\frac92\rangle$ and $|\frac92,
-\frac72\rangle$.  The only other state in the ground-state manifold to
which the open channel can couple through the dominant electrostatic
potential has atoms in the states $|\frac92,-\frac92\rangle$ and
$|\frac72,-\frac72\rangle$.  Thus, $|o\rangle= |\frac92-\frac92,
\frac92-\frac72\rangle$, and the only closed channel is $|c\rangle
=|\frac92-\frac92,\frac72-\frac72\rangle$.  Note that the open and
closed channels have the state $|\frac92,-\frac92\rangle$ in common.
Thus, we can immediately conclude that any initial occupation of the state
$|\frac92,-\frac92\rangle$ will have direct effects on the closed
channel Feshbach state. Such effects  cannot be described by the
usual Bose-Fermi models which treat the Feshbach molecule as a
point Boson. It was demonstrated in Ref.\ \cite{Parish} using a schematic
mean field theory that these effects can be important for BCS-BEC cross-over
physics. Here, we will demonstrate that the occupation of the
closed channel state can have significant effects on scattering properties
using an effective theory based on a microscopic description of the atom-atom
interaction.  The threshold energy for this resonance is given by
\begin{eqnarray}
&&E_{\rm th}=E_{\frac72,-\frac72}-E_{\frac92,-\frac72}\,,
\label{Eth}\\
&&E_{\frac72,-\frac72}=-{\textstyle\frac14}\big[A +
{\sqrt{32A^2 + [ 2C -2 D-7A]^2}} + 14D\big],
\nonumber\\
&&E_{\frac92,-\frac72}=-{\textstyle\frac14}\big[A
- {\sqrt{32A^2 + [ 2 C -2 D-7 A]^2}} + 14 D\big].
\nonumber
\end{eqnarray}
With the values of $A=-1.37\cdot 10^{-2}$~K, $C=1.34\cdot
10^{-4}\,B/{\rm G}$~K, and $D=1.19\cdot 10^{-8}\,B/{\rm G}$~K
from Ref.~\cite{PethickBook}, we can Taylor expand around $B_0=201.6$\,G
to obtain
\begin{eqnarray}
E_{\rm th}\simeq0.084K+1.78\mu_B(B-B_0)\left[1+0.0188\frac{B-B_0}{100G}\right]
\label{EthTaylor}
\end{eqnarray}
which is approximately linear in $B-B_0$ for fields near the resonance.

We emphasize that our effective theory is not limited to two-channel
problems.  It is readily generalized to more than two channels if appropriate
for other atomic resonances.

The total Hamiltonian describing a system
of particles in the scattering states $|o\rangle$ and $|c\rangle$ interacting
via $V(r)$ is
\begin{gather}
\hat{H}(B)=
\sum_{{\vec{k}},\alpha}\epsilon_{\alpha,k}\hat{a}^\dagger_{{\vec{k}}\alpha}
\hat{a}_{{\vec{k}}\alpha}
+\nonumber\\
\frac{1}{\mathcal{V}}\sum_{\vec{K},{\vec{q}}}
\hat{\Psi}^\dagger_{{\vec{q}}'}({\vec{K}})
\begin{bmatrix}V_{cc}({\vec{q}}',{\vec{q}})&V_{co}({\vec{q}}',{\vec{q}})\\
V_{oc}({\vec{q}}',{\vec{q}})&V_{oo}({\vec{q}}',{\vec{q}})
\end{bmatrix}
\hat{\Psi}_{{\vec{q}}}({\vec{K}})
\label{Hamiltonian}
\end{gather}
where
$
\hat{\Psi}^\dagger_{{\vec{q}}}({\vec{K}})=[
\hat{a}^\dagger_{{\vec{K}}/2+{\vec{q}}\alpha_4}
\hat{a}_{{\vec{K}}/2-{\vec{q}}\alpha_3}^\dagger,
\hat{a}^\dagger_{{\vec{K}}/2+{\vec{q}}\alpha_2}\hat{a}_
{{\vec{K}}/2-{\vec{q}}\alpha_1}^\dagger]
$
describes the creation of an atomic pair in the closed and open channels with
center of mass momentum ${\vec{K}}$ and relative momentum ${\vec{q}}$.
Here, $\epsilon_{\alpha,k}=\epsilon_\alpha+k^2/2m$ with $m$ the mass of the
atoms, and ${\mathcal{V}}$ is the volume of the system. The interaction matrix
elements of $V(r)$ in (\ref{vcentral}) between the different scattering
channels are denoted $V_{ij}({\vec{q}}',{\vec{q}})$ with $i,j=o,c$.  They
depend on $B$ through the $B$-dependence of the states $|o\rangle$ and
$|c\rangle$.

The scattering process is described by the scattering matrix,
$T_{ij}(\omega,{\vec{K}},{\vec{q}}',{\vec{q}})$, which depends on
the center-of-mass frequency and momentum $(\omega,\vec{K})$
and the relative momenta  $\vec{q}$ and $\vec{q}'$ in
the incoming and outgoing  channels respectively.
The scattering matrix obeys in the ladder approximation the Lippmann-Schwinger equation
\begin{gather}
\begin{bmatrix}
T_{cc}&T_{co}\\
T_{oc}&T_{oo}
\end{bmatrix}^{-1}
=
\begin{bmatrix}
V_{cc}&V_{co}\\
V_{oc}&V_{oo}
\end{bmatrix}^{-1}
-
\begin{bmatrix}
G_c&0\\
0&G_o
\end{bmatrix}
\label{LippmannSchwinger}
\end{gather}
where the propagator for two atoms in the open channel is
\begin{equation}
G_{o}(\omega,{\vec{K}},{\vec{q}})=
\frac{1-f_{\alpha_1}(\frac12\vec{K}-\vec{q})-
f_{\alpha_2}(\frac12\mathbf{K}+\vec{q})}
{\omega+i\delta-\frac{K^2}{4\,m}-\frac{q^2}{m}} \, ,
\end{equation}
and
\begin{equation}
G_{c}(\omega,{\vec{K}},{\vec{q}},B)=\frac{1-f_{\alpha_3}(\frac12\vec{K}
-\vec{q})-
f_{\alpha_4}(\frac12\vec{K}+\vec{q})}
{\omega+i\delta-E_{\rm th}(B)-\frac{K^2}{4m}-\frac{q^2}{m}}
\end{equation}
is the propagator for a pair of atoms in the closed channel.  Here
$f_{\alpha}$ is the Fermi distribution function appropriate for the
hyperfine state $|\alpha\rangle$. Note that (\ref{LippmannSchwinger})
is a matrix equation both in the relative momenta $({\mathbf{q}}',
{\mathbf{q}})$ and in the channels $(i,j)$.

\subsection{Effective interaction}
We now define an  effective interaction $U$ appropriate for calculating
the low energy scattering properties of the system. This is done by
considering the Lippmann-Schwinger equation in vacuum for $\omega = 0$ and
ignoring the hyperfine splitting of the atomic levels
\begin{equation}
\begin{bmatrix}
U_{cc}&U_{co}\\
U_{oc}&U_{oo}
\end{bmatrix}^{-1}
=
\begin{bmatrix}V_{cc}&V_{co}\\
V_{oc}&V_{oo}
\end{bmatrix}^{-1}
-
\begin{bmatrix}
G^{\rm vac}&0\\
0&G^{\rm vac}
\end{bmatrix}
\label{LippmannSchwingerVac}
\end{equation}
where $G^{\rm vac}(q)=(i\delta-q^2/m)^{-1}$ is the vacuum pair propagator
ignoring the hyperfine splitting.  When the scattering is dominated by the
presence of a single Feshbach state in a closed channel, the effective
interaction mediated by the closed channel state is rank one separable: Its
Fourier transform can be expressed in terms of the product of the Fourier
transforms $\langle\phi_k|\hat{V}|\phi_{\rm m}\rangle$,
where $|\phi_{\rm m}\rangle$ is the molecular state and $|\phi_k\rangle$
is a plane wave state of the two atoms~\cite{AndyBook}.  For the low
energies relevant for dilute atomic gases, the typical wavelength $q^{-1}$
of the open channel scattering atoms is such that $qr_C\ll 1$ where $r_C$
is the characteristic length scale of $\langle {\mathbf{r}}|\hat{V}|
\phi_{\rm m}\rangle$.  For these long wavelengths, the solution of
(\ref{LippmannSchwingerVac}) can be written from (\ref{vcentral}) as
\begin{equation}
\hat{U}({\mathbf{q}}',{\mathbf{q}})=
\frac{4\pi}{m}\left[\frac{a_s+3a_t}{4}+(a_t-a_s){\mathbf{S}}_1\cdot{\mathbf{S}}_2
\right]g(q')g(q)
\label{EffectiveInteraction}
\end{equation}
where $a_s$ and $a_t$ are the scattering lengths for the singlet $V_s(r)$
and triplet $V_t(r)$ potentials, respectively.  Cut-off effects of the
potential (for $r_C>0$) are contained in the form factor $g(q)\rightarrow 0$
for $qr_C\rightarrow \infty$.  The atom-atom potential is thus characterized
by the three parameters:  $a_s$, $a_t$, and (implicitly) $r_C$.  Using
(\ref{EffectiveInteraction}), one can readily calculate the low energy
effective interaction matrix elements $\langle i|U|j\rangle$
for the scattering channels $|i\rangle$. Note that these matrix elements
depend on $B$ through the $B$-dependence of the eigenstates of (\ref{Hspin}).
For $^{40}$K, a simple calculation expressing the hyperfine states
$|9/2,-9/2\rangle$, $|9/2,-7/2\rangle$, and $|7/2,-7/2\rangle$  in terms
of the eigenstates of the electron and nuclear spin for a given $B$-field
yields
\begin{equation}
\begin{bmatrix}
U_{cc}&U_{co}\\
U_{oc}&U_{oo}
\end{bmatrix}=\frac{4\pi}{m}
\begin{bmatrix}
\frac{c_{7}a_s+a_t}{1+c_{7}}&\frac{a_t-a_s}{\sqrt{1+c_{7}}\sqrt{1+c_{9}}}\\
\frac{a_t-a_s}{\sqrt{1+c_{7}}\sqrt{1+c_{9}}}&\frac{c_{9}a_s+a_t}{1+c_{9}}
\end{bmatrix}
\label{U40K}
\end{equation}
where $c_7=(2x-7-\sqrt{81-28x+4x^2})^2/32$ and $c_9=(2x-7+
\sqrt{81-28x+4x^2})^2/32$ with $x=(C-D)/A$ incorporate the $B$ dependence
of the effective interaction.

With (\ref{EffectiveInteraction}) [or (\ref{U40K}) for $^{40}$K], we
have obtained a low energy effective interaction.  Since $U$ is a solution
to the scattering problem in vacuum, it includes the influence of the
non-resonant coupling to high momentum states in both open and closed channels.

\subsection{Scattering matrix}
We can now use $U$ as an effective low energy interaction for the exact
solution of the scattering problem.  The resonant scattering matrix can
be expressed in terms of $U$ using (\ref{LippmannSchwinger}) and
(\ref{LippmannSchwingerVac}) with the result that
\begin{equation}
\begin{bmatrix}
T_{cc}&T_{co}\\
T_{oc}&T_{oo}
\end{bmatrix}^{-1}
=
\begin{bmatrix}U_{cc}&U_{co}\\
U_{oc}&U_{oo}
\end{bmatrix}^{-1}
-
\begin{bmatrix}
\Delta G_c&0\\
0&\Delta G_o
\end{bmatrix}
\label{LippmannSchwingerEffective}
\end{equation}
where $\Delta G_{o}(\omega,{\mathbf{K}},{\mathbf{q}})=
G_{o}(\omega,{\mathbf{K}},{\mathbf{q}})-G^{\rm vac}(q)$
and $\Delta G_{c}(\omega,{\mathbf{K}},{\mathbf{q}},B)=
G_{c}(\omega,{\mathbf{K}},{\mathbf{q}},B)-G^{\rm vac}(q)$.
The pair propagators $\Delta G_{o}$ and $\Delta G_{c}$ provide corrections
to the scattering matrix arising from the hyperfine level structure which
leads to the presence of a Feshbach resonance and corrections due to the
presence of a medium in both channels. Note that since the effective
interaction $U$ includes the coupling to high energy states, our theory
as defined in (\ref{LippmannSchwingerEffective}) is properly renormalized
in the sense that it yields finite results even for $r_C=0$ corresponding to a
$\delta$-function interaction.

With $U$ given by (\ref{EffectiveInteraction}), we can readily solve
(\ref{LippmannSchwingerEffective}) to obtain the scattering matrix
in the open channel
\begin{equation}
T_{oo}^{-1}=\left(U_{oo}+\frac{U_{oc}\Pi_cU_{co}}{1-U_{cc}\Pi_c}
\right)^{-1}-\Pi_o
\label{Tsolution}
\end{equation}
where
\begin{equation}
\Pi_o(\omega,{\mathbf{K}})=\int
\frac{d^3q}{(2\pi)^3}\left[G_o(\omega,{\mathbf{K}},{\mathbf{q}})-
G^{\rm vac}(q)\right]g^2(q)
\label{Piopen}
\end{equation}
and
\begin{equation}
\Pi_c(\omega,{\mathbf{K}},B)=\int
\frac{d^3q}{(2\pi)^3}\left[G_c(\omega,{\mathbf{K}},{\mathbf{q}},B)-
G^{\rm vac}(q)\right]g^2(q).
\label{Piclosed}
\end{equation}
The integrals in (\ref{Piopen})-(\ref{Piclosed}) can performed
straightforwardly in the vacuum case for a zero range ($r_C=0$)
potential with the result that $\Pi^{\rm
vac}_o(\omega,{\mathbf{K}})=
-im^{3/2}\sqrt{\tilde{\omega}}/(4\pi)+
{\mathcal{O}}(\sqrt{m\,\tilde{\omega}}\,r_C)$ where
$\tilde{\omega}=\omega-{\mathbf{K}}^2/(4m)$ is the energy in the
center of mass frame.   Similarly, 
$\Pi^{\rm vac}_c(\omega,
{\mathbf{K}},B)=\Pi_o^{\rm vac}[\omega-E_{\rm
th}(B),{\mathbf{K}} ]+{\mathcal{O}}(\sqrt{mE_{\rm th}}r_C)$.

\subsection{Feshbach resonance}
Consider now the case when a Feshbach resonance is present at $B=B_0$.
For fields close to $B_0$, it is standard to parameterize the zero energy
vacuum scattering matrix as
\begin{equation}
T_{oo}^{vac}=\frac{4\pi a_{\rm bg}}{m}\left(1-\frac{\Delta B}{B-B_0}\right)
\label{phenomenology}
\end{equation}
where $a_{\rm bg}$ is the (non-resonant) background scattering length and
$\Delta B$ the width of the resonance.  We see from (\ref{Tsolution}) that
the zero energy ($\omega=0$) scattering matrix in a vacuum is
\begin{equation}
T^{\rm vac}(0)=U_{oo}+|U_{oc}|^2\frac{\Pi^{\rm vac}_c(0)}{1-
U_{cc}\Pi^{\rm vac}_c(0)}
\label{Tvac}
\end{equation}
since $\Pi_o^{\rm vac}(\omega=0)=0$.  A comparison of (\ref{phenomenology})
with (\ref{Tvac}) allows us to express the phenomenological low energy
scattering parameters $a_{\rm bg}$, $\Delta B$, and $B_0$ in terms of the
microscopic parameters $a_s$, $a_t$ and $r_C$ which characterize the
atom-atom interaction.  The resonant part of the scattering coming from
the coupling between the open and closed channels is clearly given by the
second term in (\ref{Tvac}), which depends on the $B$ field through $U$ and
$\Pi^{\rm vac}_c(0)$.  The position of the resonance $B_0$ is thus determined
by
\begin{equation}
1-{U_{cc}(B_0)}\Pi^{\rm vac}_c(B_0,\omega=0)=0.
\label{B0}
\end{equation}
A straightforward pole expansion in $B-B_0$ of (\ref{Tvac}) yields
\begin{eqnarray}
&&T_{\rm bg}=\frac{4\pi a_{\rm
bg}}{m}=\Bigg[U_{oo}-\frac{U_{oc}^2}{U_{cc}}
\nonumber \\
&&+\frac{2U_{oc}\partial_BU_{oc}-U_{oc}^2\frac{\partial^2_B(U_{cc}-
U_{cc}^2\Pi^{\rm vac}_c)}{2\partial_B(U_{cc}-
U_{cc}^2\Pi^{\rm vac}_c)}}{\partial_B(U_{cc}-U_{cc}^2\Pi^{\rm vac}_c)}
\Bigg]\Bigg|_{\omega=0,B=B_0}
\label{abg}
\end{eqnarray}
for the background scattering and
\begin{equation}
T_{\rm bg}\Delta B=
-\frac{U_{oc}(B_0)^2/U_{cc}(B_0)}{\left.\partial_B[1-{U_{cc}}(B)
\Pi_c^{\rm vac}(B,0)]\right|_{B_0}}
\label{DeltaB}
\end{equation}
for the width of the resonance.  With (\ref{EffectiveInteraction}),
(\ref{Tsolution}),  (\ref{B0}), (\ref{abg}), and (\ref{DeltaB}), we have
finally arrived at our low energy ($kr_C\ll 1$) effective theory.
Equation (\ref{Tsolution}) gives the full low energy/momentum dependent
multi-channel scattering matrix as a function of the parameters $a_s$ and
$a_t$ (and $r_C$) through $U$ in (\ref{EffectiveInteraction}).  Medium
effects due to the occupation states in \emph{both} open and closed channels
are included explicitly included through $\Pi_o$ and $\Pi_c$.  Furthermore,
(\ref{B0}), (\ref{abg}), and (\ref{DeltaB}) express the physical parameters
$a_{\rm bg}$, $B_0$, and $\Delta B$ describing zero energy scattering in
vacuum in terms of $a_s$ and $a_t$ and $r_C$.  Alternatively, these
equations can be used to fix $a_s$ and $a_t$ and $r_C$ from experimental data.

We emphasize that our theory is finite and well defined even for $r_C=0$
(i.e., $g(q)=1$).  The form factors $g(q)$ have been retained in order
to describe finite range effects. These can important since the hyperfine
splitting energy $E_{\rm th}$ can be comparable to $\hbar^2/(mr_C^2)$.

\section{Magnetic moment of the Feshbach molecule}

\begin{figure}[htbp]
\begin{center}\vspace{0.0cm}
\rotatebox{0}{\hspace{-0.cm}
\resizebox{7.5cm}{!}{\includegraphics{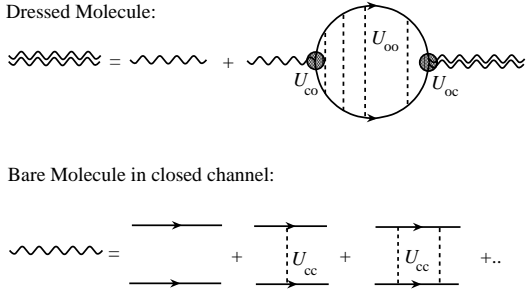}}}
\caption{The full (dressed) molecule given by  (\ref{moleculeprop})
is indicated by the double wavy lines whereas the bare molecule in the closed
channel, excluding coupling to the open channel, is given by the single wavy
line. Fermions are indicated by straight lines, and the intrachannel
couplings $U_{oo}$ and $U_{cc}$ are indicated by dashed lines. The coupling
$U_{oc}$ between the open and closed channels is indicated by $\bullet$.}
\label{MoleculeFig}
\end{center}
\end{figure}

In the spirit of Landau Fermi theory, it is often convenient to express the
full energy- and momentum-dependent multi-channel scattering matrix in
terms of physical observables only. To do this, it is necessary to introduce
one more parameter: the magnetic moment of the Feshbach molecule in a
vacuum. The energy of the molecule in a vacuum is determined by the
poles of the full pair propagator in the closed channel:
\begin{eqnarray}
D_{\rm vac}^{-1}({\mathbf{K}},\omega)&=&{\Pi_c^{\rm vac}}^{-1}-U_{cc}
-{U_{oc}}^2\frac{\Pi_o^{\rm vac}}{1-U_{oo}\Pi_o^{\rm vac}}\nonumber\\
&=&D^{-1}_{\rm bare}-{U_{oc}}^2\frac{\Pi_o^{\rm vac}}{1-U_{oo}\Pi_o^{\rm vac}}.
\label{moleculeprop}
\end{eqnarray}
The first term, $D^{-1}_{\rm bare}={\Pi_c^{\rm vac}}^{-1}-U_{cc}$, describes the
propagation of a pair of atoms interacting in the closed channel only
(i.e., the ``bare'' molecule) while the final term describes the coupling
to the open channel (``dressing''). The molecule propagator is illustrated
diagrammatically in Fig.\ \ref{MoleculeFig}.
To obtain an expression for the magnetic moment of the molecule for
fields close to $B_0$, we expand (\ref{moleculeprop}) about $B=B_0$ and
$\tilde{\omega}=0$ to obtain 
\begin{eqnarray}
z\,D_{\rm vac}^{-1}({\mathbf{K}},\omega)&\simeq&\tilde{\omega}-\Delta\mu(B-B_0)\nonumber\\
&+&i\frac{\Delta B\Delta\mu}{1/a_{\rm bg}+i\sqrt{m\tilde{\omega}}}
 \sqrt{m\tilde{\omega}}
\end{eqnarray}
where $\tilde{\omega}=\omega-K^2/4m$ is the energy in
the center of mass frame as before and
\begin{equation}
\Delta \mu=
-\frac{\partial_B[{\Pi_c^{\rm vac}}^{-1}-U_{cc}]|_{B_0}}
{\partial_\omega{\Pi_c^{\rm vac}}^{-1}-{U_{oc}}^2\partial_\omega
\Delta\Pi_o}=z\Delta\mu_{\rm bare}
\label{Deltamu}
\end{equation}
is the magnetic moment of the molecule. We have defined 
\begin{gather}
\Delta \mu_{\rm bare}=-\frac{\partial_B[{\Pi_c^{\rm vac}}^{-1}-U_{cc}]|_{B_0}}
{\partial_\omega{\Pi_c^{\rm vac}}^{-1}}\nonumber\\
\approx \partial_BE_{\rm th}+2\,E_{\rm th} U_{cc}^{-1}\partial_B U_{cc}
\label{Deltamubare}
\end{gather}
using $\Pi_c^{\rm vac}(B,\omega)\propto[E_{\rm
th}(B)-\omega]^{1/2}$  for $r_C=0$
to calculate the partial derivatives. In addition, 
\begin{eqnarray}
z^{-1}&=&1-{U_{oc}}^2
\partial_\omega\Delta\Pi_o^{\rm vac}/\partial_\omega {\Pi_c^{\rm vac}}^{-1}\nonumber\\
&=&1-T_{\rm bg}\Delta B\Delta \mu_{\rm bare}
\partial_\omega\Delta\Pi_o^{\rm vac} \, ,
\label{z}
\end{eqnarray}
where we have used $T_{\rm bg}\Delta B\Delta \mu_{\rm bare}\partial_\omega {\Pi_c^{\rm vac}}^{-1}=U_{oc}^2$
following from (\ref{DeltaB},\ref{Deltamubare}) and explicitly separated the threshold
$\sqrt{\omega}$ term in $\Pi_o$ by writing
\begin{equation}
\Pi_o^{\rm vac}(\omega)=-i\frac{m^{3/2}}{4\pi}
\sqrt{\omega}+\Delta\Pi_o(\omega).
\end{equation}
The term $\Delta\Pi_o$ contains the corrections to the pair propagator
in the open channel coming from the  finite range ($r_C\neq 0$) of the
potential. Equation (\ref{Deltamu})-(\ref{z}) provides a clear physical
interpretation of the various contributions to the magnetic moment of
the molecule. Equation (\ref{Deltamubare}) gives the magnetic moment of
the bare molecule when the interaction with atom pairs in the open channel
is ignored.  It consists of a contribution from the magnetic moment
difference between the open and closed channels, $\partial_BE_{\rm th}$ and
a contribution from the magnetic field dependence of the interaction matrix
elements in the closed channel. For the $^{40}$K resonance discussed above,
we have from (\ref{EthTaylor}) $\partial_BE_{\rm th}=1.78\mu_B$.
The second term in (\ref{Deltamubare}) can be estimated as
\begin{eqnarray}
2\,E_{\rm th}\,U_{cc}^{-1}\partial_B U_{cc}\approx 0.42\,\mu_B\,
\frac{a_s-a_t}{a_s+a_t}\left[1+0.89\,\frac{a_s-a_t}{a_s+a_t}\right]^{-1}
.
\nonumber
\end{eqnarray}
The factor $z$ given by (\ref{z}) gives the renormalization of the
magnetic moment due to coupling of the molecule to high momentum atom
pairs in the open channel (excluding the threshold effect given explicitly
by the $\sqrt{\omega}$-term).  This renormalization was first included
systematically in a low energy effective theory in Ref.\ \cite{bruunpethick}
where it was argued that it can, in general, be significant.  To see this,
we use a simple cut-off form factor $g(q)=\Theta(r_C^{-1}-q)$ in
(\ref{Piopen}) to calculate $\Delta\Pi_0$ as 
\begin{gather}
z^{-1}=1-\frac{2}{\pi}\Delta\mu_{\rm bare}\Delta Bma_{\rm bg}r_C\nonumber\\
=1-0.35\frac{\Delta\mu_{\rm bare}}{\mu_B}\frac{\Delta B}{10\rm{G}}\frac{m}{40u}
\frac{a_{\rm bg}}{100\rm{\AA}}\frac{r_C}{10\rm{\AA}}
\label{zphysical}
\end{gather}
where $1\,u$ is the atomic mass unit. For the Feshbach resonance of
$^{40}$K considered in this paper, $\Delta B\sim {\mathcal{O}}(10\rm{G})$ and
$a_{\rm bg}\sim {\mathcal{O}}(100\rm{\AA})$ (see next section).  It then
follows from (\ref{zphysical}) that the renormalization of the magnetic
moment of the molecule due to the coupling to high momentum atom pairs in
the open channel can be significant (i.e.\ $z<1$) if $r_C\gtrsim 10$\AA.
Likewise, the renormalization can be significant for the very broad resonance for $^6$Li at
$B\simeq830$G with  $\Delta B\sim 300$G~ and $a_{\rm bg}\sim
{\mathcal{O}}(1000\rm{\AA})$~\cite{Grimexp}.  However, for the narrow
$^6$Li resonance at $B_0\simeq543$G and $\Delta B=0.23$ G studied in
Ref.~\cite{Strecker}, (\ref{zphysical}) indicates that the renormalization
of the magnetic moment away from threshold described by $z$ can be ignored,
i.e.\ $z\simeq 1$.

\begin{figure}[htbp]
\begin{center}\vspace{0.0cm}
\rotatebox{0}{\hspace{-0.cm}
\resizebox{7.5cm}{!}{\includegraphics{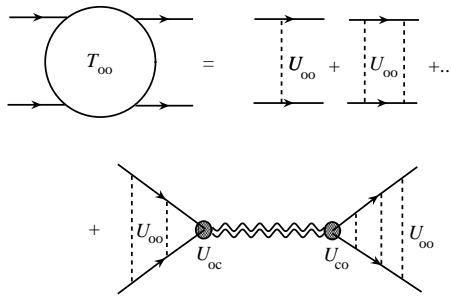}}}
\caption{The open-open channel scattering matrix split into scattering in
the open channel only (first line) and scattering via the closed channel
(second line) as given by (\ref{Tatas}).}
\label{TFig}
\end{center}
\end{figure}

Having introduced the magnetic moment of the molecule, we can now express
the full frequency- and momentum-dependent scattering matrix in a medium
in terms of the physical parameters $a_{\rm bg}$, $B_0$, $\Delta B$, and
$\Delta\mu$.  To do this, we rewrite the open-open channel scattering matrix
given by (\ref{Tsolution}) in the form
\begin{gather}
T_{oo}=\frac{U_{oo}}{1-U_{oo}\Pi_o}+\frac{U_{oc}^2/
[1-U_{oo}\Pi_o]^2}{\Pi_c^{-1}-U_{cc}
-{U_{oc}}^2\frac{\Pi_o}{1-U_{oo}\Pi_o}}\nonumber\\
=\frac{U_{oo}}{1-U_{oo}\Pi_o}+
\frac{U_{oc}}{1-U_{oo}\Pi_o}D\frac{U_{co}}{1-U_{oo}\Pi_o}
.
\label{Tatas}
\end{gather}
The first term describes the scattering via the open channel interaction
only, and the second term describes scattering via the closed channel. The
 in-medium molecule propagator $D$ is given by (\ref{moleculeprop}) with the
replacement $\Pi^{\rm vac}\rightarrow \Pi$.
This splitting of the scattering into open channel and closed channel
processes is illustrated diagrammatically in Fig.\ \ref{TFig}.
Comparison with (\ref{moleculeprop}) reveals that the second term in
(\ref{Tatas}) describes the presence of the Feshbach molecule (including
medium effects) leading to the resonance. Note that both terms in
(\ref{Tatas}) contribute to the background scattering given by (\ref{abg}).
Indeed, a Taylor expansion around $\omega=0$ and $B=B_0$ and comparison
with the vacuum expression for the scattering matrix (\ref{phenomenology})
yields 
\begin{equation}
T_{oo}=\frac{T_{\rm bg}}{\left(1+\frac{\Delta\mu\Delta B}
{\tilde{\omega}+h(\omega)-\Delta\mu(B-B_0)}\right)^{-1}
-T_{\rm bg}\Pi^{\rm inf}_o(\omega)}
\label{T(mu,abg)}
\end{equation}
where $\Pi^{\rm inf}_o(\omega,{\mathbf{K}})$ is the pair propagator in the open channel
assuming $r_C=0$.  The term
\begin{gather}
h(\omega,{\mathbf{K}})=\frac{\Pi_c(\omega,{\mathbf{K}},B)
-\Pi_c^{\rm vac}(B,0)-
\left.\partial_\omega\Pi_c^{\rm vac}(B_0,\omega)\right|_0\tilde{\omega}}
{\left.\partial_\omega\Pi_c^{\rm vac}(B_0,\omega)\right|_0+
\left(\frac{U_{oc}}{U_{cc}}\right)^2\partial_\omega
\Delta\Pi_o^{\rm vac}}\nonumber\\
=z\left[
\frac{\Pi_c(\omega,{\mathbf{K}},B)-\Pi_c^{\rm vac}(B,0)}
{\left.\partial_\omega\Pi_c^{\rm vac}(B_0,\omega)\right|_0}
-\tilde{\omega}\right]
\label{composite}
\end{gather}
in (\ref{T(mu,abg)}) describes effects coming from the fact that the
Feshbach state is a composite two-fermion object including medium effects
in the closed channel.

With (\ref{T(mu,abg)}), we have expressed the full scattering matrix
in terms of the physical parameters $a_{\rm bg}$, $B_0$,
$\Delta B$, and $\Delta\mu$.  These, in turn, can be calculated from  $a_s$
and $a_t$ with the aid of (\ref{B0})-(\ref{DeltaB}) and (\ref{Deltamu}).
Thus, (\ref{T(mu,abg)}) and (\ref{Tsolution}) are two formulations of the same
result, the former expresses the in-medium scattering matrix in terms of the
microscopic parameters characterizing the atom-atom interaction; the latter
provides a formulation valid close to the resonance in terms of the physical
observables characterizing the resonance. Our effective theory thus provides
a link between these descriptions.

\section{Medium effects}
We now consider the shift of the resonance position due to the presence
of a medium. To do this, we rewrite (\ref{T(mu,abg)}) as
\begin{gather}
T_{oo}=\frac{T_{\rm bg}}{1-T_{\rm bg}\Pi_o^{\rm inf}}\nonumber\\
+\frac{T_{\rm bg}\Delta\mu\Delta B/[1-T_{\rm bg}
\Pi_o^{\rm inf}]^2}{\tilde{\omega}+h(\omega)-\Delta\mu(B-B_0)
-T_{\rm bg}\Delta\mu\Delta B\frac{\Pi_o^{\rm inf}}{1-T_{\rm bg}
\Pi_o^{\rm inf}}}.
\label{Tmedium}
\end{gather}
We see clearly that the term $T_{\rm bg}\Delta\mu\Delta B\Pi_o^{\rm inf}$
in the denominator describes the effects of the presence of atoms in the
open channel whereas $h(\omega)$  describes the effects of the presence
of atoms in the closed channel.

We first calculate medium effects due to the presence of open channel
atoms. We see from (\ref{Piopen}) that medium effects in the open channel
shift $\Pi_o$ by
\begin{equation}
\int\frac{d^3q}{(2\pi)^3}\frac{f_{\alpha_1}(\frac12\vec{K}-\vec{q})
+f_{\alpha_2}(\frac12\mathbf{K}+\vec{q})}
{\frac{K^2}{4\,m}+\frac{q^2}{m}-\omega}=\frac{mk_F}{\pi^2}
\end{equation}
for $\vec{K}=0$ and $\omega=0$ where $k_F$ is the Fermi momentum for both
open channel hyperfine states (we have assumed equal population) and
$k_BT\ll k_F^2/2m$. The resulting shift $\delta B$ in the position of the
center of mass ${\mathbf{K}}=0$ resonance from its vacuum position $B_0$
is then readily determined from (\ref{Tmedium})
\begin{equation}
\delta B=-\frac{4}{\pi}k_Fa_{\rm bg}\Delta B.
\label{dBopen}
\end{equation}
Taking a density of $n=2\times 10^{12}cm^{-3}$,  $a_{\rm bg}=290a_0$, and
$\Delta B=8$G relevant for the $^{40}$K resonance considered here,
(\ref{dBopen}) yields $\delta B=-0.61$G for the shift of the
position of the resonance due to the occupation of states in the open channel.
Of course, in a typical radio-frequency (RF) experiment which probes the
scattering of pairs of atoms within the Fermi sea with varying center of
mass momenta ${\mathbf{K}}$~\cite{Grimexp}, it is necessary to average
(\ref{Piopen}) and (\ref{Tmedium}) appropriately over momenta in order
to obtain the average medium shift of the resonance position.

The effects due to the population of the closed channel states can likewise
be calculated from $h(\omega)$.  The medium shift of $h(\omega)$ is obtained
from (\ref{composite}) and (\ref{Piclosed}) as
\begin{equation}
\int\frac{d^3q}{(2\pi)^3}\frac{f_{\alpha_3}(\frac12\vec{K}
-\vec{q})+f_{\alpha_4}(\frac12\mathbf{K}+\vec{q})}
{E_{\rm th}+\frac{K^2}{4\,m}+\frac{q^2}{m}
-\omega}\simeq\frac{n_c}{E_{\rm th}} \, ,
\end{equation}
where $n_c$ is the density of closed channel atoms and $E_{\rm th}\gg
(3\pi^2n_C)^{2/3}/2m$ by assumption.  The shift of the resonance position
due to the occupation of closed channel states follows from (\ref{Tmedium}) as
\begin{equation}
\delta B=-\frac{8\pi}{m^{3/2}}\frac{n_c}{\sqrt{E_{\rm th}}
\Delta\mu_{\rm bare}}=-\frac{2n_cU_{cc}}{\Delta\mu_{\rm bare}}.
\label{dBclosed}
\end{equation}
For the $^{40}$K resonance, an assumed total density of  $n=2\times
10^{12}\,{\rm cm}^{-3}$ corresponds to $n_c=10^{12}\,{\rm cm}^{-3}$
since the open and the closed channels share one hyperfine state.
Using $E_{\rm th}=0.084K$ from (\ref{EthTaylor}) and assuming
$\Delta\mu_{\rm bare}\sim \mu_B$ yields $\delta B=-1.7\times10^{-6}$\,G
for the shift of the position of the resonance due to the occupation of
states in the closed channel. The shift $\delta B$ is very small because 
$E_{\rm th}\gg \epsilon_F$ for this resonance where $\epsilon_F$ is the Fermi energy.

The medium effects are treated within the ladder approximation in the 
present paper. Close to resonance when interaction effects are strong,
it is not obvious that this approximation is adequate.
However, the good agreement 
between the single channel thermodynamic calculations based on the 
ladder approximation 
and Monte Carlo results indicate that the ladder approximation
includes the most important physics, even close to resonance~\cite{Pieri}.

\section{Comparison with Bose-Fermi effective theories}
A comparison with other recent low energy effective theories for atomic
Feshbach resonances is now in order. Most of these theories are based on
a Bose-Fermi model where the presence of the Feshbach molecule is put in
by hand as a point boson~\cite{Feshbachrefs,bruunpethick}.  In the present
approach, on the other hand, the Feshbach molecule emerges dynamically and
is treated correctly as a composite two-fermion object consisting of atom
pairs in both the open and the closed channels.  This allows for the
description of effects coming from the two-fermion nature of the molecule
through $\Pi_c$ in (\ref{Tsolution}) or the term $h(\omega)$ in
(\ref{T(mu,abg)})-(\ref{composite}).  The simpler Bose-Fermi theory, which
describes the Feshbach state as a point boson whose propagator has linear
$\omega$-dependence, is recovered only when $h(\omega)=0$.  Furthermore,
the present theory provides expressions for all low energy scattering and
bound state properties, including medium effects in both open \emph{and}
closed channels, through the parameters $a_s$, $a_t$, and $r_C$ characterizing
the atom-atom interaction and thus provides a nice link between a microscopic
(coupled-channels) description of the scattering and more phenomenological
descriptions.  This includes a physically clear picture of the various
contributions to the magnetic moment of the Feshbach molecule.  Once $a_s$,
$a_t$, and $r_C$ are fixed, our approach affords a systematic description of
Feshbach resonances in \emph{all} the relevant channels.

\section{Rethermalization Rate}
Finally, we analyze a recent experiment on the Feshbach resonance at
$B_0\simeq 201.6$ G for $^{40}$K atoms in the hyperfine states $|9/2,
-9/2\rangle$ and $|9/2,-7/2\rangle$.  In refs.\,\cite{Loftus,Regal}, the
elastic collision rate for $^{40}$K atoms was measured near this Feshbach
resonance using the following rethermalization technique:  The gas was
``heated'' preferentially in one spatial direction, and the relaxation toward
equilibrium with a uniform temperature in all directions was then followed
in the time evolution of the rms cloud radii.  The rethermalization rate
was then extracted from an exponential fit of the aspect ratio as a
function of time.

A variational expression for the relaxation time of temperature anisotropies
can be given~\cite{PethickBook} as
\begin{equation}
\frac{1}{\tau_T}=
\frac{\langle\langle\Phi_T\Gamma[\Phi_T]\rangle\rangle}{\langle\langle
\Phi_T^2\rangle\rangle}.
\label{ThermalRate}
\end{equation}
Here,
\begin{equation}
\langle\langle\ldots\rangle\rangle=\int d^3r\int\frac{d^3p}{(2\pi)^3}
\ldots f_0(1-f_0)
\end{equation}
where $f_0$ is the equilibrium distribution function for $|9/2,-9/2\rangle$
and $|9/2,-7/2\rangle$ atoms and where we treat the trap potential $V_{\rm
trap}({\mathbf{r}})$ using the Thomas-Fermi approximation with a spatially
dependent chemical potential $\mu({\mathbf{r}})=\mu-V_{\rm trap}
({\mathbf{r}})$. The populations of $|9/2,-9/2\rangle$ and $|9/2,-7/2\rangle$
atoms are assumed to be equal. The deviation function appropriate for
thermal anisotropy is
\begin{equation}
\Phi_T=p_z^2-p^2/3+\frac{m^2}{3}(2\omega_z^2z^2-\omega_{\perp}^2r_{\perp}^2) \ ,
\end{equation}
where $\omega_z$ and $\omega_\perp$ are the trapping frequencies in the axial and transverse
directions (we assume an axially symmetric trap). The linearized collision  operator is
\begin{gather}
f_0(1-f_0)\Gamma[\Phi]({\mathbf{r}},{\mathbf{p}})=
\int\frac{d^3p_1}{(2\pi\hbar)^3}\int
d\Omega\frac{d\sigma}{d\Omega}
|{\mathbf{v}}-{\mathbf{v}}_1|\nonumber\\
[\Phi+\Phi_1-\Phi'-\Phi_1']f^0f^0_1(1-{f^0}')(1-{f^0_1}'),\label{collintegral}
\end{gather}
where $d\sigma/d\Omega$ is the on-shell ($\omega=p^2/2m+p_1^2/2m$)
differential cross section and $\Omega$ is the solid angle for the
direction of the relative outgoing momentum
${\mathbf{p}}_r'=({\mathbf{p}}'-{\mathbf{p}}_1')/2$ with respect
to the relative incoming momentum
${\mathbf{p}}_r=({\mathbf{p}}-{\mathbf{p}}_1)/2$~\cite{PethickBook}.
The calculation of the integrals in (\ref{ThermalRate}) proceeds as in
Ref.~\cite{Kavoulakis}.  In the classical limit, these integrals
can be evaluated analytically yielding
\begin{equation}
\frac{1}{\tau_T}=\frac{4}{5N}\left(\frac{k_BT}{\pi m}\right)^{1/2}
\langle\sigma\rangle \int d^3 r n({\mathbf{r}})^2
\label{ThermalRateClass}
\end{equation}
where $N$ is the total number of trapped atoms (in both hyperfine states),
$n$ the total density of atoms, and
\begin{equation}
\langle\sigma\rangle=\frac{1}{3}\int dx
x^7\sigma(\sqrt{mk_BT}x)e^{-x^2}\,.
\label{SigmaAverage}
\end{equation}
The factor of $x^7$ in (\ref{SigmaAverage}) comes from the
weight function $\Phi_T$ appropriate for temperature relaxation.
Note that (\ref{ThermalRateClass}) differs from the averaged
cross section arising from the collision relaxation time.
The latter results from a thermal average of the cross section
weighted with the relative velocity only, which corresponds
to the factor $x^3$ in (\ref{SigmaAverage}).

\begin{figure}
\includegraphics[clip=true,width=8cm]{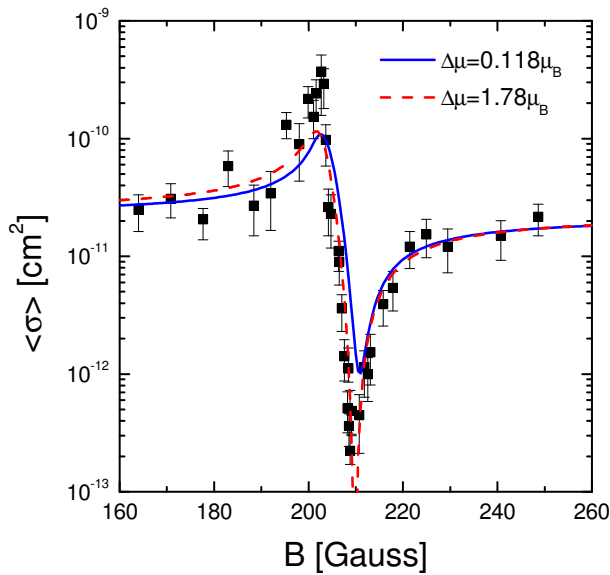}
\caption{Fit to the experimental rethermalization cross section~\cite{Regal}
with (\ref{SigmaAverage})-(\ref{Tvacuum}) for two values of the molecule
magnetic moment at $T=4.4~\mu$K.  The fit parameters are given in
(\ref{smallmupar}) for $\Delta\mu=0.118~\mu_B$ and (\ref{largemupar})
for $\Delta\mu=1.78~\mu_B$.}
\label{fig:Fit}
\end{figure}

We now use $\sigma=m^2|T_{oo}|^2/4\pi$ with $T_{oo}$ given by
(\ref{T(mu,abg)}) to analyze the experimental data.  In
Refs.~\cite{Loftus,Regal}, the experiments are performed at
temperatures $T\sim 2 T_{\rm F}$ and densities $n\sim
10^{12}$~cm$^{-3}$\,. We can therefore use the classical results
(\ref{ThermalRateClass}) and (\ref{SigmaAverage}).
Likewise, $T_{oo}$ in (\ref{T(mu,abg)}) can be approximated by the
vacuum on-shell scattering matrix
\begin{gather}
T_{oo}=
\frac{\frac{4\,\pi}{m}\,a_{\rm bg}}{\left(1
+\frac{\Delta\mu\Delta B}{q^2/m-\Delta\mu(B-B_0)}\right)^{-1}+ia_{\rm bg}q}
\label{Tvacuum}
\end{gather}
where ${\mathbf{q}}$ is the relative momentum of the two scattering atoms.
At present, knowledge of the experimental value of the magnetic moment of the
Feshbach molecule is limited.  We therefore consider two cases of a large
and  a small value of $\Delta\mu$\,.  We take $\Delta\mu=
\partial_BE_{\rm th}=1.78~\mu_B$ from (\ref{EthTaylor}) as a large value
and $\Delta\mu=\Delta\mu_0=0.118~\mu_B$ from Ref.~\cite{bruunpethick} as
a small value.  Using (\ref{SigmaAverage})-(\ref{Tvacuum}), we can now
analyze the experimental results~\cite{Loftus,Regal}.  For the two values
of $\Delta\mu$ above, we tune the parameters $\Delta B$ and $a_{\rm bg}$
keeping the resonance position fixed at $B_0=201.6$~G.
The results of our fits are presented in Fig.~\ref{fig:Fit}.
For $\Delta\mu=0.118\,\mu_B$ we find
\begin{eqnarray}
\Delta B=6.5~{\rm G} \ \ \ \ {\rm and}\ \ \ \ a_{\rm bg}=286.7\,a_0\,.
\label{smallmupar}
\end{eqnarray}
The fit with $\Delta\mu=1.78\,\mu_B$ gives
\begin{eqnarray}
\Delta B=8.1~{\rm G} \ \ \ \ {\rm and}\ \ \ \ a_{\rm bg}=296.7\,a_0\,.
\label{largemupar}
\end{eqnarray}
 In Ref.~\cite{Regal}, the authors give
the fitted values of the singlet and triplet scattering lengths as
$a_s\approx 104.8\, a_0$ and $a_t\approx 174\, a_0$\,.  Substituting
these values in (\ref{abg}), we find $a_{\rm bg}\simeq 170\,a_0$, which
is close to the value $164\,a_0$ reported in Ref.~\cite{Bohn}. 
This value differs somewhat from ours ($a_{\rm bg}\simeq290a_0$).
The reason for this discrepancy can be the experimental uncertainty 
in determining the number of atoms trapped, and/or the fact that the authors 
in  Ref.~\cite{Regal} use a different thermal averaging procedure 
than that given by  (\ref{SigmaAverage})~\cite{BohnUnpublished}.

\begin{figure}
\includegraphics[clip=true,width=8cm]{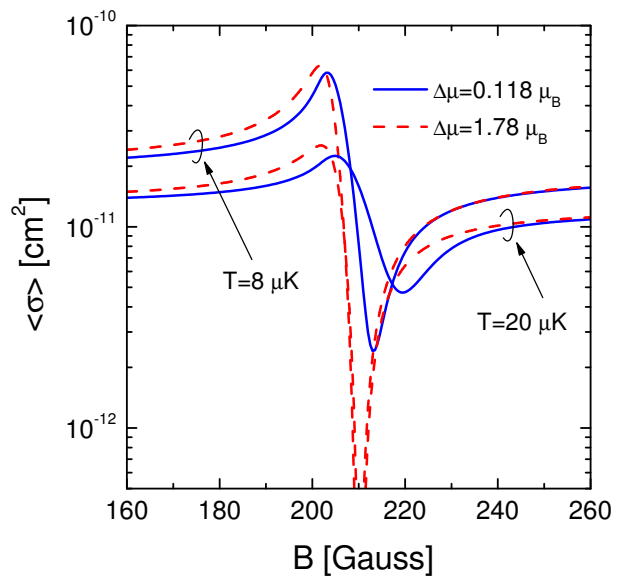}
\caption{The rethermalization cross section (\ref{SigmaAverage}) with
(\ref{Tvacuum}) for two values of the molecule
magnetic moment and the temperatures $T=8~\mu$K and $20~\mu$K.
The fit parameters are given in
(\ref{smallmupar}) for $\Delta\mu=0.118~\mu_B$ and
(\ref{largemupar}) for $\Delta\mu=1.78~\mu_B$.}
\label{fig:Temp}
\end{figure}

The fit parameters (\ref{smallmupar})-(\ref{largemupar}) do not depend strongly on the
value of the molecule magnetic moment at the temperature $T=4.4~\mu$K.
This is because $\Delta\mu\Delta B\gg k_BT$ for both values of $\Delta\mu$:
$\Delta\mu\Delta B/k_B=55\mu K$ for $\Delta\mu=0.118\,\mu_B$ and
$\Delta\mu\Delta B/k_B=980\mu K$ for $\Delta\mu=1.78\,\mu_B$. In fact,
the overall shape of the rethermalization curve as a function of $B-B_0$ 
(e.g.\ the position of the resonance and the magnitude of the maximum) 
becomes sensitive to $\Delta\mu$ only when
\begin{equation}
\Delta\mu\Delta B\sim k_BT.
\end{equation}
To illustrate this, we show in Fig.~\ref{fig:Temp} the cross section
(\ref{SigmaAverage}) evaluated for temperatures $T=8~\mu$K and $T=20~\mu$K
and for the two values of $\Delta \mu$ above.  We see a clear difference
in the shape of the curve depending on the temperature. 

\begin{figure}
\includegraphics[clip=true,width=8cm]{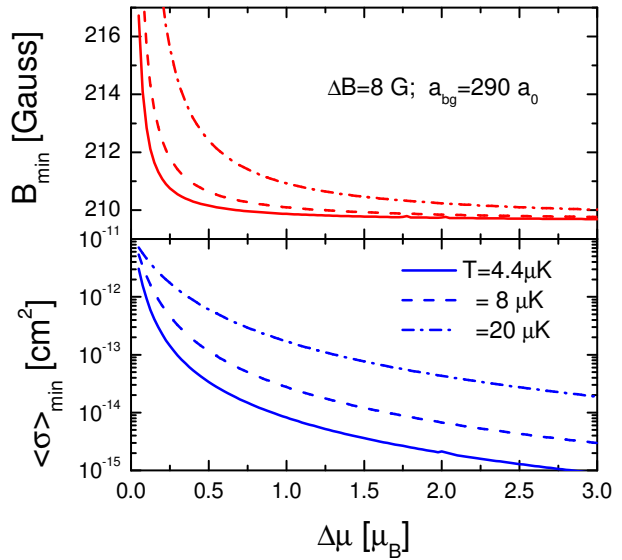}
\caption{The position $B-B_0=B_{\rm min}$ and magnitude $\langle\sigma_{\rm min}\rangle$ 
of the minimal value of the rethermalization cross section as a function of the molecule
magnetic moment $\Delta\mu$ for  $T=4.4~\mu$K, $T=8~\mu$K, and $T=20~\mu$K.}
\label{fig:Minimum}
\end{figure}

The \emph{magnitude} of the minimum of the  rethermalization rate located at $B-B_0\sim\Delta B$ (for 
$k_BT\ll \Delta\mu\Delta B$) is however sensitive to the temperature
even for $k_BT\ll \Delta\mu\Delta B$. This can easily be understood as follows. 
If one ignores the $q^2/m$ term in (\ref{Tvacuum}), the
cross section is identically zero for $B-B_0=\Delta B$ for all
${\mathbf{q}}$ leading to a vanishing rethermalization rate. But from 
$q^2/m\sim k_BT$ is follows from (\ref{Tvacuum}) that the minimum of the 
rethermalization rate at $B-B_0\sim\Delta B$ scales as $k_BT/\Delta\mu\Delta B$
for $k_BT\ll \Delta\mu\Delta B$. Thus, the minimum of the rethermalization rate is 
sensitive to the temperature or the value of the molecule magnetic moment even for 
$k_BT\ll \Delta\mu\Delta B$. To illustrate this, we plot in Fig.~\ref{fig:Minimum}
the position $B_{\rm min}=B-B_0$ and the magnitude $\langle\sigma_{\rm min}\rangle$ 
of the minimal value of the rethermalization cross 
section as a function of the molecule magnetic moment $\Delta\mu$ for  $T=4.4~\mu$K, $T=8~\mu$K, 
and $T=20~\mu$K.
We see that the position of the minimum does not depend on $\Delta\mu$ when $k_BT\ll \Delta\mu\Delta B$
whereas the magnitude of the minimum value of the rethermalization rate does. 

The temperature dependence of the rethermalization rate could be used to determine 
the magnetic moment of the Feshbach molecule: One
can measure this magnetic moment by fitting the measured rethermalization
cross section as a function of temperature using the theory described above.
A scan of temperatures from $k_BT\ll \Delta\mu\Delta B$ to $k_BT\gg \Delta\mu\Delta B$ 
is likely to yield a fairly accurate value of $\Delta\mu$. For $k_BT\ll \Delta\mu\Delta B$,
the Feshbach resonance gives rise to a sharp maximum and minimum  in the rethermalization 
scattering cross section whereas for $k_BT\gg \Delta\mu\Delta B$, the rethermalization 
cross section is a much smoother function of the $B$ field. 
Furthermore, if one is able to measure the magnitude
 of the minimum rethermalization rate which is sensitive to the temperature for all $T$, 
the size of the magnetic moment can even be determined for $k_BT\ll \Delta\mu\Delta B$.

\section{Effective Range}
We now demonstrate that the energy-dependent, Feshbach-mediated interaction
introduces a new length scale: the effective range $r_{\rm eff}$. The
effective range expansion for the vacuum scattering matrix is given
by~\cite{LandauLifshitz}
\begin{equation}
T_{oo}=\frac{4\pi a}{m}\frac{1}{1-\frac{1}{2}q^2ar_{\rm eff}+i\, qa}
\label{TeffectiveRange}
\end{equation}
where $a=a_{\rm bg}[1-\Delta B/(B-B_0)]$ is the resonant scattering length.
This expansion can be obtained from (\ref{Tvacuum}) in the limit $|B-B_0|
\ll \Delta B $ with the effective range given as
\begin{equation}
r_{\rm eff}=-\frac{2}{m\Delta\mu\Delta Ba_{\rm bg}}.
\label{reff}
\end{equation}
The $q^2/m$ term in (\ref{Tvacuum}) coming from the momentum dependence
of the Feshbach molecule energy can therefore introduces a new length
scale: An effective range given by (\ref{reff}).  Equation (\ref{reff}) was
also derived in Ref.~\cite{BruunUnivarsality} in the context of a Bose-Fermi
model.

Note that the effective range given by (\ref{reff}) is negative as a
direct consequence of the energy dependence of the multi-channel
interaction.  In a recent paper, finite range effects on the physics of
cold atomic gases in the BEC-BCS crossover regime was examined using a
single channel model~\cite{Parish2}.  The atom-atom potential used in
Ref.~\cite{Parish2} has a positive effective range and thus does not
provide a correct description of finite range effects for resonant
multi-channel atom-atom scattering.  Similarly, the effective range of
the single channel interaction used in a recent Monte-Carlo study also has
$r_{\rm eff}\ge 0$~\cite{Carlson}.  In general, it is often difficult
and sometimes impossible to obtain a negative effective range in single
channel models of resonant scattering.  This feature emerges naturally
in multi-channel models.

It is common to neglect the effective range term by adopting a simple
approximate form of the unitarized scattering matrix
\begin{equation}
T=\frac{4\pi a}{m}\frac{1}{1+iqa}.
\label{Tunitary}
\end{equation}
This approximation leads to the idea of universality.  When $|a| \rightarrow
\infty$, the only remaining length scale in the problem is the interparticle
spacing, and the properties of the gas should therefore be independent of
the microscopic details of the atom-atom interaction. The physics of
interacting gases in the universal limit has been examined by a number of
authors~\cite{Carlson,Heiselberg,Ho,Baker}.  However, we have now
demonstrated explicitly that there is an additional length scale in the
problem given by (\ref{reff}). This length scale can be neglected only
when calculating properties with a typical wave length $q$ for which $qr_{\rm eff}\ll 1$;
the thermodynamic properties of the gas are independent of $r_{\rm eff}$ only when 
\begin{equation}
|r_{\rm eff}|\ll r_{12}
\end{equation}
where $r_{12}$ is the average interparticle spacing. For low temperatures where $q\sim k_F$, 
using (\ref{reff}), this condition can  be written as
\begin{equation}
|\Delta\mu\Delta B|\gg \frac{\epsilon_F}{k_F|a_{\rm bg}|}.
\label{UnitarityCondition}
\end{equation}
The simple unitarized form for the scattering given by (\ref{Tunitary})
therefore holds only for broad resonances as defined by
(\ref{UnitarityCondition}).  In particular, one can expect universal behavior
only for such broad resonances~\cite{BruunUnivarsality}.

For the  Feshbach resonance for $^{40}$K at $B_0\simeq201.6$G, we find
$\Delta B\sim 10$G and $a_{\rm bg}\sim 290a_0$ as explained above.  Assuming $\Delta \mu\sim \mu_B$ and
 $\epsilon_F=10^{-6}$K, this yields
$\Delta\mu\Delta B\gg \epsilon_F/k_Fa_{\rm bg}$ and one would expect the gas to exhibit
universal behavior close to resonance.  This is also true for the very
broad resonance for $^6$Li at $B \simeq 830$G with $\Delta B \sim 300$G.
However, for the narrow $^6$Li resonance at $B_0 \simeq 543$G, $\Delta B=0.23$G, and  
$a_{\rm bg}\sim 80a_0$~\cite{Strecker}, 
we find $\epsilon_F\sim 3k_Fa_{\rm bg}\mu_B \Delta B$ for $\epsilon_F=10^{-6}$K. 
One can therefore not expect the effective range to be an irrelevant length scale
for this narrow resonance.

\section{Conclusions}
A low energy effective theory for the in-medium scattering of alkali atoms
was obtained using a microscopic multi-channel description of the atom-atom
interaction.  Expressions for all scattering properties in terms of a
few parameters characterizing the atom-atom potential were derived thereby
providing a link between a microscopic multi-channel description of Feshbach
resonances and more phenomenological approaches. The presence of a Feshbach
resonance emerged naturally from the theory as a two-particle bound state,
and the energy dependence of the Feshbach-mediated interaction was shown to
introduce a negative effective range inversely proportional to the width of the
resonance. Furthermore, medium effects arising from the occupation of both
open and closed channel states were calculated.  The theory was shown to
allow the explicit calculation of corrections to commonly used approximations
such as the neglect of the effective range and the treatment of the Feshbach
molecule as a point boson. We also analyzed a recent rethermalization
experiment on $^{40}$K atoms and showed that measurement of the
rethermalization rate as a function of temperature would permit the
determination of the magnetic moment of the Feshbach molecule.
\\

The authors acknowledge useful discussions with  J.\ L.\ Bohn and C.\ J.\ Pethick.
The work of E.E.K.\ was supported in part by the US Department of Energy 
under contract No.\ DE-FG02-87ER40328.


\begin{thebibliography}{99}
\bibitem{Experiments}C.\ A.\ Regal, M.\ Greiner, and D.\ S.\ Jin, Phys.\
Rev.\ Lett.\ \textbf{92}, 040403 (2004);
M.\ W.\ Zwierlein \textit{et al}., \textit{ibid} \textbf{92}, 120403 (2004);
 T.\ Bourdel \textit{et al}.,  \textit{ibid}  \textbf{93}, 050401 (2004);
J.\ Kinast \textit{et al}.,  \textit{ibid}  \textbf{92}, 150402 (2004);
C.\ Chin \textit{et al}., Science \textbf{305}, 1128 (2004).
\bibitem{2body} See e.g.\ D.\ J.\ Heinzen in \textit{Bose-Einstein Condensation in Atomic Gases},
Proceedings of the Enrico Fermi International School of Physics, Vol.\ CXL, ed.\ M.\ Ignuscio, S.\ Stringari, and C.\ E.\ Wieman,
(IOS Press, Amsterdam, 1999).
\bibitem{Randeria} See e.g.\ P.\  Pieri, L.\ Pisani, G.\ C.\ Strinati,
Phys. Rev. B \textbf {70}, 094508  (2004) and references therein.
\bibitem{Feshbachrefs}M.\ Holland \textit{et al}., Phys.\ Rev.\ Lett.\ \textbf{87}, 120406 (2001);
S.\ J.\ J.\ M.\ F.\ Kokkelmans \textit{et al}., Phys. Rev. A \textbf {65}, 053617  (2002);
Y.\ Ohashi and A.\ Griffin, \textit{ibid}  \textbf{67}, 033603 (2003);
R.\ Duine and H.\ T.\ C.\ Stoof, J.\ Opt.\ B \textbf{5}, S212 (2003).
\bibitem{bruunpethick} G.\ M.\ Bruun and C.\ J.\ Pethick,  Phys.\ Rev.\ Lett.\ \textbf{92}, 140404  (2004).
\bibitem{PethickBook} See e.g.\ C.\ J.\ Pethick and H.\ Smith,
\textit{Bose-Einstein Condensation in Dilute Gases} (Cambridge University Press, Cambridge 2002).
\bibitem{Loftus} T.\ Loftus \textit{et al}., Phys.\ Rev.\ Lett.\ \textbf{88}, 173201  (2002).
\bibitem{Regal}C.\ A.\ Regal, C.\ Ticknor, J.\ L.\ Bohn, and D.\ S.\ Jin,
Phys.\ Rev.\ Lett.\ \textbf{90}, 053201 (2003).
\bibitem{AndyBook} See e.g.\ G.\ E.\ Brown  and A.\ D.\ Jackson,
\textit{The nucleon-nucleon interaction} (North-Holland,  1976).
\bibitem{Parish} M.\ M.\ Parish, B.\ Mihaila, B.\ D.\ Simons, and P.\ B.\ Littlewood,
 cond-mat/0409756.
\bibitem{Strecker} K.\ E.\ Strecker, G.\ B.\ Partridge, and R.\ Hulet, Phys.\ Rev.\ Lett.\ \textbf{91}, 080406  (2003).
\bibitem{Grimexp} M.\ Bartenstein \textit{et al}., Phys.\ Rev.\ Lett.\ \textbf{94}, 103201  (2005).
\bibitem{Pieri} A.\ Perali, P.\ Pieri, and G.\ C.\ Strinati, cond-mat/0501631.
\bibitem{Kavoulakis} G.\ M.\ Kavoulakis, C.\ J.\ Pethick, and H.\ Smith, Phys.\ Rev.\ A
\textbf {61}, 053603  (2000).
\bibitem{Bohn} J.\ L.\ Bohn,  Phys.\ Rev.\ A \ \textbf{61}, 053409 (2000).
\bibitem{BohnUnpublished}  J.\ L.\ Bohn (private communication).
\bibitem{LandauLifshitz} See e.g.\ L.\ D.\ Landau and E.\ M.\ Lifshitz,
\textit{Quantum Mechanics} (Pergamon Press, 1977).
\bibitem{BruunUnivarsality} G.\ M.\ Bruun, Phys. Rev. A \textbf {70}, 053602  (2004).
\bibitem{Parish2} M.\ Parish \textit{et al}.,   Phys.\ Rev.\ B \textbf {71}, 064513  (2005).
\bibitem{Carlson} J.\ Carlson, S-Y Chang, V.\ R.\ Pandharipande, and K.\ E.\ Schmidt,
Phys.\ Rev.\ Lett.\ \textbf{91} 050401.
\bibitem{Ho} T.\ L.\ Ho and E.\ J.\ Mueller, Phys.\ Rev.\ Lett.\ \textbf{92}, 160404 (2004);
T.\ L.\ Ho, Phys.\ Rev.\ Lett.\ \textbf{92}, 090402 (2004).
\bibitem{Heiselberg} H.\ Heiselberg, Phys.\ Rev.\ A \textbf{63}, 043606 (2001)
\bibitem{Baker} G.\ A.\ Baker, Phys.\ Rev.\ C \textbf{60}, 054311 (1999).
\end{thebibliography}
\end{document}